# A General Simulation Framework for Supply Chain Modeling: State of the Art and Case Study

Antonio Cimino[1], Francesco Longo[2] and Giovanni Mirabelli[3]

[1] Mechanical Department, University of Calabria,
Rende (CS), 87036, Italy

[2] Mechanical Department, University of Calabria,
Rende (CS), 87036, Italy

[3] Mechanical Department, University of Calabria,
Rende (CS), 87036, Italy

## Abstract

Nowadays there is a large availability of discrete event simulation software that can be easily used in different domains: from industry to supply chain, from healthcare to business management, from training to complex systems design. Simulation engines of commercial discrete event simulation software use specific rules and logics for simulation time and events management. Difficulties and limitations come up when commercial discrete event simulation software are used for modeling complex real world-systems (i.e. supply chains, industrial plants). The objective of this paper is twofold: first a state of the art on commercial discrete event simulation software and an overview on discrete event simulation models development by using general purpose programming languages are presented; then a Supply Chain Order Performance Simulator (SCOPS, developed in C++) for investigating the inventory management problem along the supply chain under different supply chain scenarios is proposed to readers.

**Keywords:** *Discrete Event Simulation, Simulation languages, Supply Chain, Inventory Management.*

## 1. Introduction

As reported in [1], discrete-event simulation software selection could be an exceeding difficult task especially for inexpert users. Simulation software selection problem was already known many years ago. A simulation buyer's guide that identifies possible features to consider in simulation software selection is proposed in [2]. The guide includes in the analysis considerations several aspects such as Input, Processing, Output, Environment, Vendor and Costs. A survey on users' requirements about discrete-event simulation software is presented in [3]. The analysis shows that simulation software with good visualization/animation properties are easier to use but limited in case of complex and non-standard problems. Further limitations include lack of software compatibility, output analysis tools, advanced programming languages. In [4] and [5] functionalities and potentialities of different commercial discrete-event simulation software, in order to support users in software selection, are reported. In this case the author provides the reader with information about software vendor, primary software applications, hardware platform requirements, simulation animation, support, training and pricing.

Needless to say that Modeling & Simulation should be used when analytical approaches do not succeed in identifying proper solutions for analyzing complex systems (i.e. supply chains, industrial plants, etc.). For many of these systems, simulation models must be: (i) flexible and parametric (for supporting scenarios evaluation) (ii) time efficient (even in correspondence of very complex real-world systems) and (iii) repetitive in their architectures for scalability purposes [6].

Let us consider the traditional modeling approach proposed by two commercial discrete event simulation software, Em-Plant by Siemens PLM Software solutions and Anylogic by Xj-Technologies. Both of them propose a typical object oriented modeling approach. Each discrete event simulation model is made up by system state variables, entities and attributes, lists processing, activities and delays. Usually complex systems involve high numbers of resources and entities flowing within the simulation model. The time required for executing a simulation run depends on the numbers of entities in the simulation model: the higher is the number of entities the higher is the time required for executing a simulation run. In addition, libraries objects, which should be used for modeling static entities, very often fall short of recreating the real system with satisfactory accuracy. In other words, the traditional modeling approach (proposed by eM-Plant and Anylogic as well as by a number of discrete event simulation software), presents two problems: (i) difficulties in modeling complex scenarios; (ii) too many entities could cause computational heavy simulation models. Further information on discrete event simulation software can be found in [7].

An alternative to commercial discrete event simulation software is to develop simulation models based on general purpose programming languages (i.e. C++, Java). The use of general purpose programming languages allows to develop ad-hoc simulation models with class-objects able to recreate carefully the behavior of the real world system.

The objective of this paper is twofold: first a state of the art on commercial discrete event simulation software and an overview on discrete event simulation models development by using general purpose programming languages are presented; then a Supply Chain Order Performance Simulator (SCOPS, developed in C++) for investigating the inventory management problem along the supply chain under different supply chain scenarios is proposed to readers.

Before getting into details of the work, in the sequel a brief overview of paper sections is reported. Section 2 provides the reader with a detailed description of different commercial discrete event simulation software. Section 3 presents a general overview of programming languages and describes the main steps to develop a simulation model based on general purpose programming languages. Section 4 presents a three stages supply chain simulation model (called SCOPS) used for investigating inventory problems along the supply chain. Section 5 describes the simulation experiments carried out by using the simulation model. Finally the last section reports conclusions and research activities still on going.

## 2. Discrete Event Simulation Software

Table 1 reports the results of a survey on the most widely used discrete event simulation software (conducted on 100 people working in the simulation field). The survey considers among others some critical aspects such as domains of application (specifically manufacturing and logistics), 3D and virtual reality potentialities, simulation languages, prices, etc. For each aspect and for each software the survey reports a score between 0 and 10. Table 1 help modelers in discrete event simulation software selection. Moreover the following sections reports a brief description of all the software of table 1 in terms of domains of applicability, types of libraries (i.e. modeling libraries, optimization libraries, etc.), input-output functionalities, animation functionalities, etc.

### 2.1 Anylogic

Anylogic is a Java based simulation software, by XJ Technologies [8], used for forecasting and strategic planning, processes analysis and optimization, optimal operational management, processes visualization. It is widely used in logistics, supply chains, manufacturing, healthcare, consumer markets, project management, business processes and military. Anylogic supports Agent Based, Discrete Event and System Dynamics modeling and simulation. The latest Anylogic version (Anylogic 6) has been released in 2007, it supports both graphical and flow-chart modeling and provides the user with Java code for simulation models extension. For input data analysis, Anylogic provides the user with Stat-Fit (a simulation support software by Geer Mountain Software Corp.) for distributions fitting and statistics analysis. Output analysis functionalities are provided by different types of datasets, charts and histograms (including export function to text files or excel spreadsheet). Finally simulation optimization is performed by using Optquest, an optimization tool integrated in Anylogic.

### 2.2 Arena

Arena is a simulation software by Rockwell Corporation [9] and it is used in different application domains: from manufacturing to supply chain (including logistics, warehousing and distribution) from customers service and strategies to internal business processes. Arena (as Anylogic) provides the user with objects libraries for systems modeling and with a domain-specific simulation language, SIMAN [10]. Simulation optimizations are carried out by using Optquest. Arena includes three modules respectively called Arena Input Analyzer (for distributions fitting), Arena Output Analyzer (for simulation output analysis) and Arena Process Analyzer (for simulation experiments design). Moreover Arena also provides the users animation at run time as well as it allows to import CAD drawings to enhance animation capabilities.

Table 1: Survey on most widely used Simulation software

|  | *Anylogic* | *Arena* | *AutoMod* | *Emplant* | *Promodel* | *Flexsim* | *Witness* |
|---|---|---|---|---|---|---|---|
| Logistic | 6.5 | 7.5 | 7 | 7.2 | 6.5 | 7 | 7.5 |
| Manufacturing | 6.6 | 7.5 | 6.5 | 7.2 | 6.7 | 6.7 | 7.5 |
| 3D Virtual Reality | 6.6 | 6.9 | 7.3 | 6.8 | 6.7 | 7.2 | 7 |
| Simulation Engine | 7 | 8 | 7.5 | 8 | 7 | 7.5 | 8 |
| User Ability | 7 | 8 | 6 | 7 | 9 | 7.5 | 8 |
| User Community | 6.2 | 9 | 6.7 | 6.5 | 7.5 | 6.6 | 8.5 |
| Simulation Language | 6.8 | 7 | 6.25 | 6.5 | 6.5 | 6.7 | 6.5 |
| Runtime | 7.5 | 7 | 6.5 | 6.5 | 7.5 | 6 | 7 |
| Analysis tools | 6.5 | 8 | 6.9 | 7.1 | 7.7 | 6 | 7.8 |
| Internal Programming | 7.2 | 7 | 6 | 7 | 6.2 | 7 | 6.5 |
| Modular Construction | 6.1 | 7 | 6 | 6.5 | 7.5 | 7 | 7 |
| Price | 7 | 6 | 5.6 | 5.8 | 7 | 5.7 | 6 |

## 2.3 Automod

Automod is a discrete event simulation software, developed by Applied Materials Inc. [11] and it is based on the domain-specific simulation language Automod. Typical domains of application are manufacturing, supply chain, warehousing and distribution, automotive, airports and semiconductor. It is strongly focused on transportation systems including objects such as conveyor, Path Mover, Power & Free, Kinematic, Train Conveyor, AS/RS, Bridge Crane, Tank & Pipe (each one customizable by the user). For input data analysis, experimental design and simulation output analysis, Automod provides the user with AutoStat [12]. Moreover the software includes different modules such as AutoView devoted to support simulation animation with AVI formats.

## 2.4 Em-Plant

Em-plant is a Siemens PLM Software solutions [13], developed for strategic production decisions. EM-Plant enables users to create well-structured, hierarchical models of production facilities, lines and processes. Em-Plant object-oriented architecture and modeling capabilities allow users to create and maintain complex systems, including advanced control mechanisms. The Application Object Libraries support the user in modeling complex scenarios in short time. Furthermore EM-Plant provides the user with a number of mathematical analysis and statistics functions for input distribution fitting and single or multi-level factor analysis, histograms, charts, bottleneck analyzer and Gantt diagram. Experiments Design functionalities (with Experiments Manager) are also provided. Simulation optimization is carried out by using Genetic Algorithms and Artificial Neural Networks.

## 2.5 Promodel

Promodel is a discrete event simulation software developed by Promodel Corporation [14] and it is used in different application domains: manufacturing, warehousing, logistics and other operational and strategic situations. Promodel enables users to build computer models of real situations and experiment with scenarios to find the best solution. The software provides the users with an easy to use interface for creating models graphically. Real systems randomness and variability can be either recreated by utilizing over 20 statistical distribution types or directly importing users' data. Data can be directly imported and exported with Microsoft Excel and simulation optimizations are carried out by using SimRunner or OptQuest. Moreover, the software technology allows the users to create customized front-and back-end interfaces that communicate directly with ProModel.

## 2.6 Flexsim

Flexsim is developed by Flexsim Software Products [15] and allows to model, analyze, visualize, and optimize any kind of real process - from manufacturing to supply chains. The software can be interfaced with common spreadsheet and database applications to import and export data. Moreover, Flexsim's powerful 3D graphics allow in-model charts and graphs to dynamically display output statistics. The tool Flexsim Chart gives the possibility to analyze the simulation results and simulation optimizations can be performed by using both Optquest as well as a built-in experimenter tool. Finally, in addition to the previous described software, Flexsim allow to create own classes, libraries, GUIs, or applications.

## 2.7 Witness

Witness is developed by Lanner Group Limited [16]. It allows to represent real world processes in a dynamic animated computer model and then experiment with "what-if" alternative scenarios to identify the optimal solution. The software can be easily linked with the most common spreadsheet, database and CAD files. The simulation optimization is performed by the Witness Optimizer tool that can be used with any Witness model. Finally the software provides the user with a scenario manager tool for the analysis of the simulation results.

## 3. General Purpose and Specific Simulation Programming Languages

There are many programming languages, general purpose or domain-specific simulation language (DSL) that can be used for simulation models development. General purpose languages are usually adopted when the programming logics cannot be easily expressed in GUI-based systems or when simulation results are more important than advanced animation/visualization [17]. Simulation models can be developed both by using discrete-event simulation software and general purpose languages, such as C++ or Java [18].

As reported in [1] a simulation study requires a number of different steps; it starts with problem formulation and passes through different and iterative steps: conceptual model definition, data collection, simulation model implementation, verification, validation and accreditation, simulation experiments, simulation results analysis, documentation and reports. Simulation model development by using general purpose programming languages (i.e. C++) requires a deep knowledge of the logical foundation of discrete event simulation. Among different aspects to be considered, it is important to underline that discrete event simulation model consists of entities, resources control elements and operations [19]. Dynamic entities flow in the simulation model (i.e. parts in a manufacturing system, products in a supply chain, etc.). Static entities usually work as resources (a system part that provides services to dynamic entities). Control elements (such as variables, boolean expressions, specific programming code, etc.) support simulation model states control. Finally, operations represent all the actions generated by the flow of dynamic entities within the simulation model. During its life within the simulation model, an entity changes its state different times. There are five different entity states [19]: Ready state (the entity is ready to be processed), Active state (the entity is currently being processed), Time-delayed state (the entity is delayed until a predetermined simulation time), Condition-delayed state (the entity is delayed until a specific condition will be solved) and Dormant state (in this case the condition solution that frees that entity is managed by the modeler). Entity management is supported by different lists, each one corresponding to an entity state: the CEL, (Current Event List for active state entity), the FEL (Future Event List for Time-delayed entities), the DL (Delay List for condition-delayed entities) and UML (User-Managed Lists for dormant entities). In particular, Siman and GPSS/H call the CEL list CEC list (Current Events Chain), while ProModel language calls it AL (Action List). The FEL is called FEP (Future Events Heap) and FEC (Future Event Chain) respectively by Siman and GPSS/H. After entities states definition and lists creation, the next step is the implementation of the phases of a simulation run: the Initialization Phase (IP), the Entity Movement Phases (EMP) and the Clock Update Phase (CUP). A detailed explanation of the simulation run anatomy is reported in [19].

## 4. A Supply Chain Simulation Model developed in C++

According to the idea to implement simulation models based on general purpose programming languages, the authors propose a three stage supply chain simulation model implemented by using the Borland C++ Builder to compile the code (further information on Borland C++ Builder can be found in [20]). The acronym of the simulation model is SCOPS (Supply-Chain Order Performance Simulator). SCOPS investigates the inventory management problem along a three stages supply chain and allows the user to test different scenarios in terms of demand intensity, demand variability and lead times. Note that such problem can be also investigated by using discrete event simulation software [21], [22], [23] and [24].

The supply chain conceptual model includes suppliers, distribution centers, stores and final customers. In the supply chain conceptual model a single network node can be considered as store, distribution center or supplier. A supply chain begins with one or more suppliers and ends with one or more stores. Usually stores satisfy final customers' demand, distribution centers satisfy stores demand and plants satisfy distribution centers demand. By using these three types of nodes we can model a general supply chain (also including more than three stages).

Suppliers, distribution centers and stores work 6 days per week, 8 hours per day. Stores receive orders from customers. An order can be completely or partially satisfied. At the end of each day, on the basis of an Order-Point, Order-Up-to-Level (s, S) inventory control policy, the stores decide whether place an order to the distribution centers or not. Similarly distribution centers place orders to suppliers according to the same inventory control policies. Distribution centers select suppliers according to their lead times (that includes production times and transportation times).

According to the Order-Point, Order-Up-to-Level policy [25], an order is emitted whenever the available quantity drops to the order point (*s*) or lower. A variable replenishment quantity is ordered to raise the available quantity to the order-up-to-level (*S*). For each item the order point *s* is the safety stock calculated as standard deviation of the lead-time demand, the order-up to level *S* is the maximum number of items that can be stored in the warehouse space assigned to the item type considered. For the *i-th* item, the evaluation of the replenishment quantity, $Q_i(t)$, has to take into consideration the quantity available (in terms of inventory position) and the order-up-to-level *S*. The inventory position (equation 1) is the on-hand inventory, plus the quantity already on order, minus the quantity to be shipped. The calculation of $s_i(t)$ requires the evaluation of the demand over the lead time. The lead time demand of the *i-th* item (see equation 2), is evaluated by using the moving average methodology. Both at stores and distribution centers levels, managers know their peak and off-peak periods, and they usually use that knowledge to correct manually future estimates based on moving average methodology. They also correct their future estimates based on trucks capacity and suppliers quantity discounts. Finally equations 3 and 4 respectively express the order condition and calculate the replenishment quantity.

$$P_i(t) = Oh_i(t) + Or_i(t) - Sh_i(t) \qquad (1)$$

$$Dlt_i(t) = \sum_{k=t+1}^{t+LT_i} Df_i(k) \qquad (2)$$

$$P_i(t) < (s_i(t) = SS_i(t)) \qquad (3)$$

$$Q_i(t) = S_i - P_i(t) \qquad (4)$$

where,
$P_i(t)$, inventory position of the *i-th* item;
$Oh_i(t)$, on-hand inventory of the *i-th* item;
$Or_i(t)$, quantity already on order of the *i-th* item;
$Sh_i(t)$, quantity to be shipped of the *i-th* item;
$Dlt_i(t)$, lead time demand of the *i-th* item;

$Df_i(t)$, demand forecast of the *i-th* item (evaluated by means of the moving average methodology);
$LT_i$, lead time of the *i-th* item;
$s_i(t)$, order point at time *t* of the *i-th* item;
$S_i$, order-up-to-level of the *i-th* item;
$SS_i(t)$, safety stock at time *t* of the *i-th* item;
$Q_i(t)$, quantity to be ordered at time *t* of the *i-th* item.

### 4.1 Supply Chain Orders Perfomance Simulator

SCOPS translates the supply chain conceptual model recreating the complex and high stochastic environment of a real supply chain. For each type of product, customers' demand to stores is assumed to be Poisson with independent arrival processes (in relation to product types). Quantity required at stores is based on triangular distributions with different levels of intensity and variability. Partially satisfied orders are recorded at stores and distribution center levels for performance measures calculation.

In our application example fifty stores, three distribution center, ten suppliers and thirty different items define the supply chain scenario. Figure 1 shows the SCOPS user interface. The SCOPS graphic interface provides the user with many commands as, for instance, simulation time length, start, stop and reset buttons, a check box for unique simulation experiments (that should be used for resetting the random number generator in order to compare different scenarios under the same conditions), supply chain configurations (number of items, stores, distribution centers, suppliers, input data, etc.). For each supply chain node a button allows to access the following information number of orders, arrival times, ordered quantities, received quantities, waiting times, fill rates. SCOPS graphic interface also allows the user to export simulation results on txt and excel files. One of the most important features of SCOPS is the flexibility in terms of scenarios definition. The graphic interface gives to the user the possibility to carry out a number of different what-if analysis by changing supply chain configuration and input parameters (i.e. inventory policies, demand forecast methods, demand intensity and variability, lead times, inter-arrival times, number of items, number of stores, distribution centers and plants, number of supply chain echelons, etc.). Figure 2 display several SCOPS windows the user can use for setting supply chain configuration and input parameters.

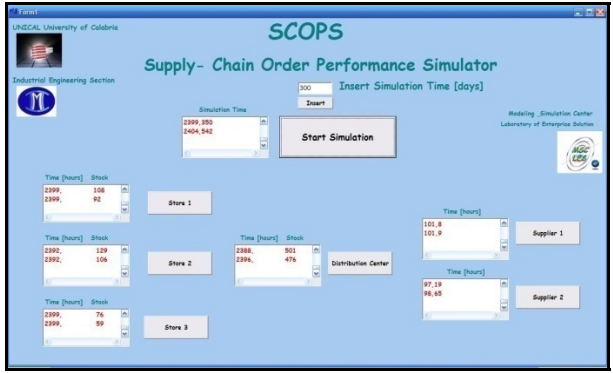

Fig. 1 SCOPS User Interface.

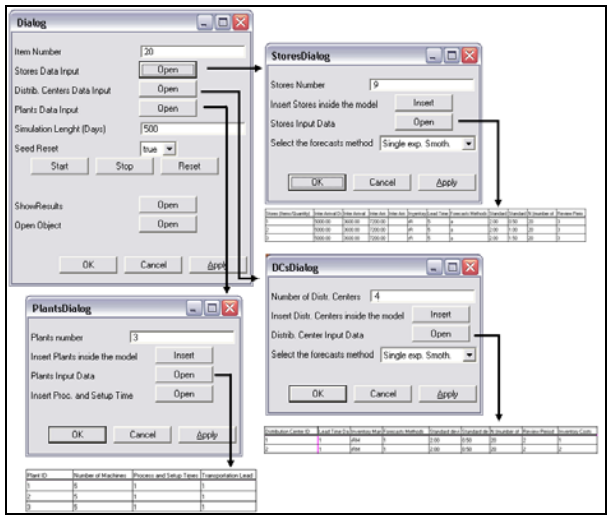

Fig. 2 SCOPS Windows.

## 4.2 SCOPS verification, simulation run length and validation

Verification and validation processes assess the accuracy and the quality throughout a simulation study [26]. Verification and Validation are defined by the American Department of Defence Directive 5000.59 as follows: verification is the process of determining that a model implementation accurately represents the developer's conceptual description and specifications, while validation is the process of determining the degree to which a model is an accurate representation of the real world from the perspective of the intended use of the model.

The simulator verification has been carried out by using the debugging technique. The debugging technique is an iterative process whose purpose is to uncover errors or misconceptions that cause the model's failure and to define and carry out the model changes that correct the errors [1]. In this regards, during the simulation model development, the authors tried to find the existence of errors (bugs). The causes of each bug has been correctly identified and the model has opportunely been modified and tested (once again) for ensuring errors elimination as well as for detecting new errors.

Before going into details of simulation model validation, it is important to evaluate the optimal simulation run length. Note that the supply chain is a non-terminating system and one of the priority objectives of such type of system is the evaluation of the simulation run length [1]. Information regarding the length of a simulation run is used for the validation. The length is the correct trade-off between results accuracy and time required for executing the simulation runs. The run length has been correctly determined using the mean square pure error analysis (MSPE). After the MSPE analysis, the simulation run length chosen is 390 days.

Choosing for each simulation run the length evaluated by means of MSPE analysis (390 days) the validation phase has been conducted by using the Face Validation (informal technique). For each retailer and each distribution centre the simulation results, in terms of fill rate, have been compared with real results. Note that during the validation process the simulation model works under identical input conditions of the real supply chain. The Face Validation results have been analyzed by several experts; their analysis revealed that, in its domain of application, the simulation model recreates with satisfactory accuracy the real system.

## 5. Supply Chain Configuration and Design of Simulation Experiments

The authors propose as application example the investigation of 27 different supply chain scenarios. In particular simulation experiments take into account three different levels for demand intensity, demand variability and lead times (minimum, medium and maximum respectively indicated with "-", "0" and "+" signs). Table 1 reports (as example) factors and levels for one of the thirty items considered and table 3 reports scenarios description in terms of simulation experiments. Each simulation run has been replicated three times (totally 81 replications).

Table 2: Factors and levels

|  | Minimum | Medium | High |
|---|---|---|---|

| Demand Intensity [inter-arrival time] | 3 | 5 | 8 |
|---|---|---|---|
| Demand Variability [item] | [18,22] | [16,24] | [14,26] |
| Lead Time [days] | 2 | 3 | 4 |

After the definition of factors levels and scenarios, the next step is the performance measures definition. SCOPS includes, among others, two fill rate performance measures defined as (i) the ratio between the number of satisfied Orders and the total number of orders; (ii) the ratio between the lost quantity and the total ordered quantity.

Simulation results, for each supply chain node and for each factors levels combination, are expressed in terms of average fill rate (intended as ratio between the number of satisfied Orders and the total number of orders).

Table 3: Simulation experiments and supply chain scenarios

| Run | Demand Intensity | Demand Variability | Lead Time |
|---|---|---|---|
| 1 | - | - | - |
| 2 | - | - | 0 |
| 3 | - | - | + |
| 4 | - | 0 | - |
| 5 | - | 0 | 0 |
| 6 | - | 0 | + |
| 7 | - | + | - |
| 8 | - | + | 0 |
| 9 | - | + | + |
| 10 | 0 | - | - |
| 11 | 0 | - | 0 |
| 12 | 0 | - | + |
| 13 | 0 | 0 | - |
| 14 | 0 | 0 | 0 |
| 15 | 0 | 0 | + |
| 16 | 0 | + | - |
| 17 | 0 | + | 0 |
| 18 | 0 | + | + |
| 19 | + | - | - |
| 20 | + | - | 0 |
| 21 | + | - | + |
| 22 | + | 0 | - |
| 23 | + | 0 | 0 |
| 24 | + | 0 | + |
| 25 | + | + | - |
| 26 | + | + | 0 |
| 27 | + | + | + |
| 1 | - | - | - |

## 5.1 Supply Chain Scenarios analysis and comparison

The huge quantity of simulation results allows the analysis of a comprehensive set of supply chain operative scenarios. Let us consider the simulation results regarding the store #1; we have considered three different scenarios (low, medium and high lead times) and, within each scenario, the effects of demand variability and demand intensity are investigated.

Figure 2 shows the fill rate trend at store #1 in the case of low lead time.

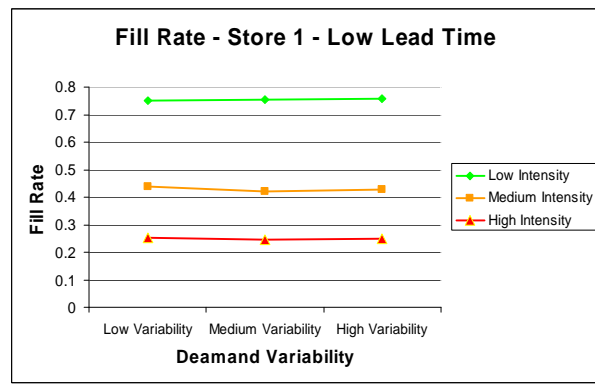

Fig. 2  Fill rate at store #1, low lead time.

The major effect is due to changes in demand intensity: as soon as the demand intensity increases there is a strong reduction of the fill rate. A similar trend can be observed in the case of medium and high lead time (figure 3 and figure 4, respectively).

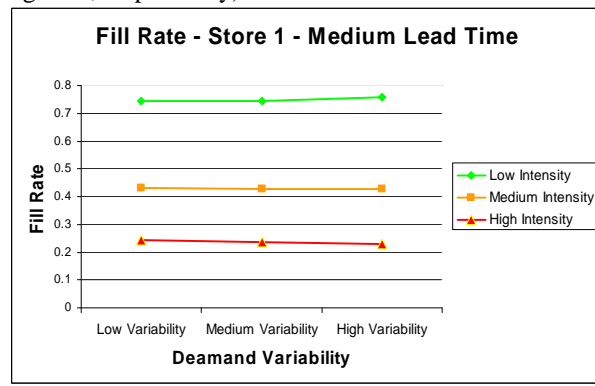

Fig. 3  Fill Rate at store # 1, medium lead time.

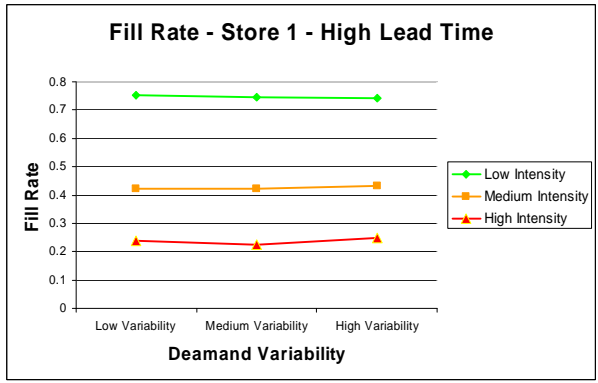

Fig. 4  Fill Rate at store # 1, high lead time.

The simultaneous comparison of figures 2, 3 and 4 shows the effect of different lead times on the average fill rate. The only minor issue is a small fill rate reduction passing from 2 days lead time to 3 and 4 days lead time.

As additional aspect (not shown in figures 2, 3, and 4), the higher is the demand intensity the higher is the average on hand inventory. Similarly the higher is the demand variability the higher is the average on hand inventory. In effect, the demand forecast usually overestimates the ordered quantity in case of high demand intensity and variability.

# 6. Conclusions

The paper first presents an overview on the most widely used discrete event simulation software in terms of domains of applicability, types of libraries (i.e. modeling libraries, optimization libraries, etc.), input-output functionalities, animation functionalities, etc. In the second part the paper proposes, as alternative to discrete event simulation software, the use of general purpose programming languages and provides the reader with a brief description about how a discrete event simulation model works.

As application example the authors propose a supply chain simulation model (SCOPS) developed in C++. SCOPS is a flexible simulator used for investigating different the inventory management problem along a three stages supply chain. SCOPS simulator is currently used for reverse logistics problems in the large scale retail supply chain.

## Acknowledgments

All the authors gratefully thank Professor A. G. Bruzzone (University of Genoa) for his valuable support on this manuscript.

**Antonio Cimino** took his degree in Management Engineering, summa cum Laude, in September 2007 from the University of Calabria. He is currently PhD student at the Mechanical Department of University of Calabria. He has published more than 20 papers on international journals and conferences. His research activities concern the integration of ergonomic standards, work measurement techniques, artificial intelligence techniques and Modeling & Simulation tools for the effective workplace design.

**Francesco Longo** received his Ph.D. in Mechanical Engineering from University of Calabria in January 2006. He is currently Assistant Professor at the Mechanical Department of University of Calabria and Director of the Modelling & Simulation Center – Laboratory of Enterprise Solutions (MSC-LES). He has published more than 80 papers on international journals and conferences. His research interests include Modeling & Simulation tools for training procedures in complex environment, supply chain management and security. He is Associate Editor of the "Simulation: Transaction of the society for Modeling & Simulation International". For the same journal he is Guest Editor of the special issue on Advances of Modeling & Simulation in Supply Chain and Industry. He is Guest Editor of the "International Journal of Simulation and Process Modelling", special issue on Industry and Supply Chain: Technical, Economic and Environmental Sustainability. He is Editor in Chief of the SCS M&S Newsletter and he works as reviewer for different international journals.

**Giovanni Mirabelli** is currently Assistant Professor at the Mechanical Department of University of Calabria. He has published more than 60 papers on international journals and conferences. His research interests include ergonomics, methods and time measurement in manufacturing systems, production systems maintenance and reliability, quality.